\documentclass[a4paper]{article}

\usepackage{INTERSPEECH2020}
\usepackage{amssymb,amsmath,epsfig}
\usepackage{textcomp}
\usepackage{amsfonts,amsthm,amsopn}
\usepackage{hyperref}
\usepackage{url}
\usepackage[capitalise]{cleveref}
\usepackage{graphicx,caption,lipsum}
\usepackage{booktabs,multirow}
\usepackage{placeins}
\usepackage{algpseudocode,algorithm}
\usepackage{color}
\usepackage[table,xcdraw]{xcolor}
\usepackage{bbm}
\usepackage{enumerate}   

\usepackage[normalem]{ulem}

\ninept
\def\reg{{\rm\ooalign{\hfil
      \raise.07ex\hbox{\scriptsize R}\hfil\crcr\mathhexbox20D}}}

\title{Open-set Short Utterance Forensic Speaker Verification \\
using Teacher-Student Network with Explicit Inductive Bias}

\makeatletter
\def\name#1{\gdef\@name{#1\\}}
\makeatother 
\name{{Mufan Sang, Wei Xia, John H.L. Hansen} }

\address{
Center for Robust Speech Systems, University of Texas at Dallas, TX 75080 
{\small \tt}}
\email{\small \tt \{mufan.sang, wei.xia, john.hansen\}@utdallas.edu}

\newcommand{\vct}[1]{\boldsymbol{\mathbf{#1}}} 

\graphicspath{{./image/}}

\begin{document}

\maketitle
\begin{abstract}
In forensic applications, it is very common that only small naturalistic datasets consisting of short utterances in complex or unknown acoustic environments are available.
In this study, we propose a pipeline solution to improve speaker verification on a small actual forensic field dataset. By  leveraging large-scale out-of-domain datasets, 
a knowledge distillation based objective function is proposed for teacher-student learning, which is applied for short utterance forensic speaker verification. 
The objective function collectively considers speaker classification loss, Kullback-Leibler divergence, and similarity of embeddings. In order to advance the trained deep speaker embedding network to be 
robust for a small target dataset, we introduce a novel strategy to fine-tune the pre-trained student model towards a forensic target domain by utilizing the model as a fine-tuning start point and a reference in regularization.
The proposed approaches are evaluated on the $1^{\rm{st}}$48-UTD forensic corpus, a newly established naturalistic dataset of actual homicide investigations consisting of short utterances recorded in uncontrolled conditions. 
We show that the proposed objective function can efficiently improve the performance of teacher-student learning on short utterances and that our fine-tuning strategy outperforms the commonly used weight decay method by providing an explicit inductive bias towards the pre-trained model.
\end{abstract}
\noindent\textbf{Index Terms}: Text-independent speaker verification, short utterance, teacher-student learning, transfer learning

\vspace{-0.5ex}
\section{Introduction}
\label{sec:intro}
Speaker verification (SV) is defined as the process of identifying the true characteristics of a speaker and to accept or discard the identity claimed by the speaker. In recent years, speaker verification has seen significant improvement with fast development of deep learning and great success of various advanced neural networks applied to deep speaker embedding systems~\cite{snyder2016deep,zhang2017end,li2017deep,wan2018generalized,snyder2018x,xia2019cross}. Generally, speaker embedding systems consist of a front-end frame-level feature extractor, an utterance-level encoding layer, one or more fully-connected layers, and an output classifier. Like the i-Vector system~\cite{dehak2010front}, speaker embedding networks can encode variable length utterances into a fix-length speaker representation. Although studies~\cite{zhang2017end,wan2018generalized} have shown that deep speaker embedding systems outperformed the traditional i-Vector solution especially on short utterances, short utterance speaker verification is still a challenging task because of insufficient phonetic information provided~\cite{kanagasundaram2011vector}. 

Recently, deep neural network (DNN) and convolutional neural network (CNN) based speaker embedding systems have been applied to solve this problem and obtain effective performance improvements~\cite{xie2019utterance,jung2019short,hajavi2019deep,gusev2020deep}. In~\cite{jung2019short}, a raw waveform CNN-LSTM architecture was proposed to extract phonetic-level features which can help compensate for missing phonetic information. In~\cite{xie2019utterance,hajavi2019deep}, some new aggregation methods such as NetVLAD layer and Time-Distributed Voting (TDV) were designed to improve the efficiency of the aggregation process. In order to maintain strong speaker discrimination, most studies use large amounts of in-domain data to train the speaker embedding networks for short utterances and evaluate them in the same domain.
However, it usually happens that only a very small size dataset is available for the specific target domain, and directly training the model from scratch on it will greatly degrade the performance. As one application of speaker verification, forensic related problems are complex and challenging because there are actual constraints on availability of speaker data and length of utterances in both test and enrollment~\cite{mandasari2011evaluation,poddar2017speaker}. Also, inconsistencies usually exist in forensic relevant datasets, such as unknown recording locations, noise, reverberation, very limited duration of useful human speech, etc.~\cite{al2017enhanced,machado2019forensic}. These variants can lead to significant performance degradation for speaker verification systems.

\begin{figure*}[th]
  \centering
    \vspace{-2mm}
    \includegraphics[width=14cm,height=5.3cm]{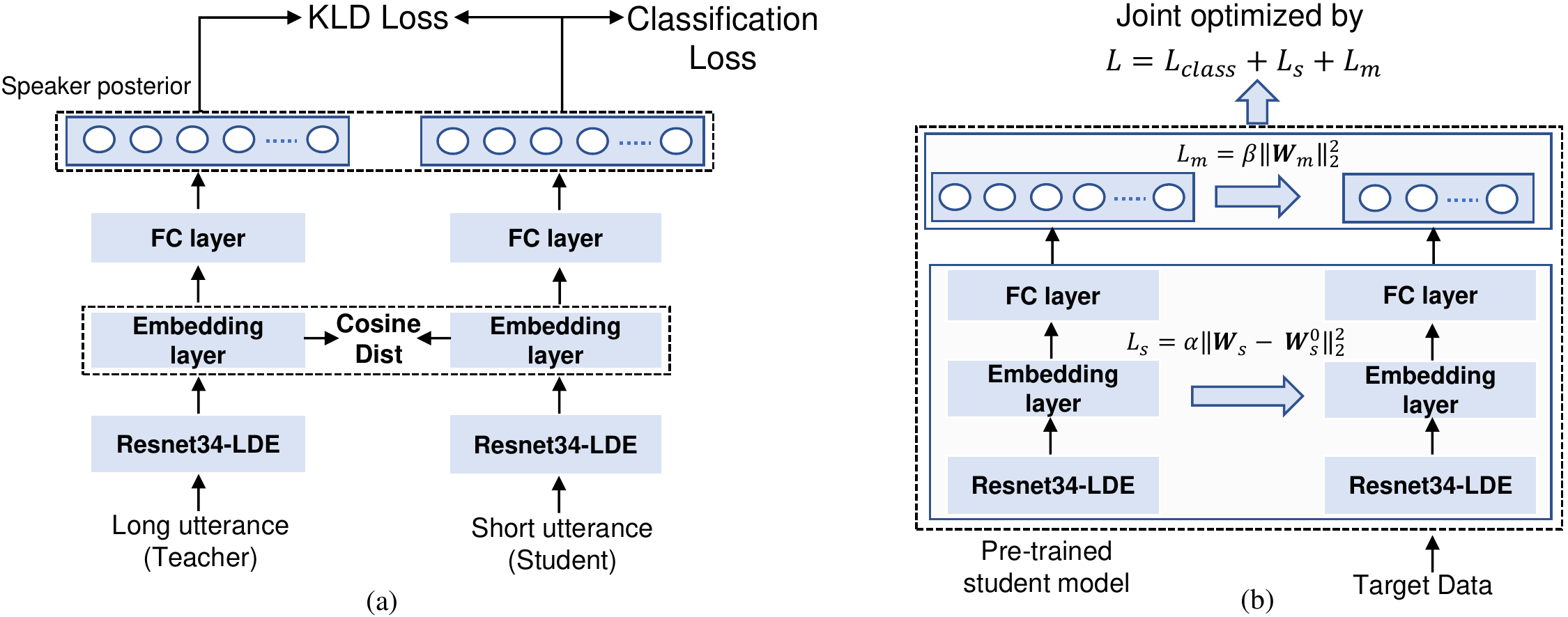}
    \vspace{-3.1mm}
  \caption{ (a) Flow-diagram of our teacher-student learning framework. The KL-divergence loss between posteriors, cosine distance between embeddings, and speaker classification loss are shown in the figure. 
  (b) Proposed fine-tuning strategy to transfer the pre-trained student model to the small target dataset} 
    \vspace{-3mm}
\end{figure*}

In this work, we collect a new challenging naturalistic dataset for forensic analysis of audio from various cities across the USA where detectives were investigating actual homicides. We focus on the problem of short utterance speaker verification in forensic domains with only a small naturalistic target dataset available. In order to alleviate this problem, first, we propose a knowledge distillation~\cite{hinton2015distilling} based objective function which collectively considers Kullback-Leibler divergence (KLD) of posteriors, similarity of embeddings, and speaker classification loss to improve performance of short utterance speaker verification with teacher-student (T-S) learning applied. Second, to avoid losing initial speaker discriminative knowledge learned from a large number of speakers from out-of-domain datasets, we introduce a new fine-tuning strategy that helps encode an explicit inductive bias towards the pre-trained model by using the pre-trained network as a fine-tuning start point as well as a reference penalized in the regularization. In this way, the fine-tuned model takes advantage of the strong speaker discriminative power on short utterances and learns new speaker characteristics from the small target dataset.

We evaluate the performance of our fine-tuning strategy based on the $L^2$ regularization (known as weight decay), which often results in a suboptimal solution for the target domain with an implicit inductive bias towards the pre-trained model. In this paper, we show that the proposed knowledge distillation loss function is able to transfer knowledge from teacher to student more effectively, and the introduced fine-tuning strategy outperforms the $L^2$ regularization by adequately preserving the learned features from large-scale source datasets.

In Sec. 2, we describe the proposed pipeline, including the teacher-student learning framework used for short utterance speaker verification and the proposed fine-tuning strategy. Data description, experiment setting, and analysis results are reported in Sec. 3 and 4. Finally, we conclude this paper in Sec. 5.


\section{Methodology}
\label{sec:method}
In this section, we introduce our proposed approaches for short utterance speaker verification on the small forensic target dataset, 
where utterances in the target domain have much shorter length as well as several variation factors compared with the source domain. 
First, the speaker embedding model should learn a strong speaker discriminative power on short utterances from large out-of-domain datasets. Next, the model is refined to perform well on short 
utterances in the target domain with only a much smaller in-domain dataset available. We utilize T-S learning with the proposed objective function to improve system performance on short utterances. Then, we use our fine-tuning strategy to adapt the trained student model from the large source dataset to the small target dataset with an explicit inductive bias.
The architecture of the entire pipeline is illustrated in Figure 1.

\subsection{Teacher-Student Learning for short utterances}

In this study, the teacher model is trained independently, and the student model is learned to mimic the probability distribution of the teacher’s output and account for speaker classification loss. Our teacher-student learning framework is described in the following sections. 

\subsubsection{Speaker classification loss}
For both teacher and student models, softmax cross-entropy loss is used as the speaker classification loss. We also explore the angular softmax (A-softmax)~\cite{liu2017sphereface} loss which has a stronger discriminative ability by maximizing the angular margin between speaker embeddings.

\subsubsection{Student model training loss}

In conventional T-S learning, the student is learned to mimic the teacher by minimizing the Kullback-Leibler divergence between the teacher and student output distributions given parallel data~\cite {lu2017knowledge}. When adopting T-S learning for a short utterance problem, the KLD function can be presented as,
\begin{align}
L_{\mathrm{KLD}}=-\sum_{i=1}^{I} \sum_{n=1}^{N} P\left(y_{n} | \vct{x}_{i,T}\right) \log \left(P\left(y_{n} | \vct{x}_{i,S}\right)\right)
\end{align}
\noindent where $y_{n}$ refers to speaker $n$ and $\vct{x}_{i,T}$ and $\vct{x}_{i,S}$ refer to the $i$-th input sample of the teacher and student models. In this work, the teacher model is fed with long utterance samples, and the student is fed with the shorter crop of the same utterances.
$P\left(y_{n} | \vct{x}_{i,T}\right)$ and $P\left(y_{n} | \vct{x}_{i,S}\right)$ are the posteriors of the $i$-th sample predicted by teacher and student models. In this way, the student model is enforced to generate similar posteriors as the teacher model.

As the representation of speakers, embedding is the key to a speaker verification system. 
The problem will be effectively addressed if a model can produce more similar or even identical embeddings for short utterances vs. long utterances. According to this idea, directly constricting the similarity of embeddings between short and long utterances would be more efficient~\cite{jung2019short,wang2019knowledge}. 
It can be achieved by minimizing the distance between embeddings of short and long utterances. We use cosine distance (COS) as the distance metric,
 with the corresponding loss function formulated as,
\vspace{-2ex}
\begin{align}
L_{\mathrm{EMD}}=-\sum_{i=1}^{1} \frac{\epsilon_{T}^{i} \cdot \epsilon_{S}^{i}}{\left\|\epsilon_{T}^{i}\right\|\left\|\epsilon_{S}^{i}\right\|}
\end{align}

\noindent where $\epsilon_{T}^{i}$ and $\epsilon_{S}^{i}$ represent the embeddings predicted by the teacher and student models for the $i$-th sample.

In optimization, we use a multi-task objective function to combine the losses introduced above. $L_{\mathrm{class}}$ denotes the speaker classification loss. Three different combinations of objective functions are described below:

(1)$L_{\mathrm{class}}+L_{\mathrm{KLD}}$: On the basis of speaker classification loss, adding soft labels enables the student model to enhance its discriminative power toward that of the teacher model trained on long utterances

(2)$L_{\mathrm{class}}+L_{\mathrm{EMD}}$: Replacing the KLD loss with the embedding-based loss enables the student model to generate more similar embeddings to that of the teacher. 

(3)$L_{\mathrm{class}}+L_{\mathrm{KLD}}+L_{\mathrm{EMD}}$: Collectively combining all three can guarantee  both the speaker discriminative power and the similarity of embeddings between short and long utterances.

We compare the efficacy of trained student networks with the baseline models trained without T-S learning. We also compare the performances of student models optimized by different objective functions mentioned above.

\subsection{Explicit Inductive bias for fine-tuning}

Considering the limited source, naturalistic contexts, and several inter-speaker and intra-speaker variations contained in the short utterance forensic corpus~\cite{bimbot2004tutorial} 
, directly training the deep speaker embedding model from scratch will not ensure discriminative power of the speakers, and will also contribute to severe overfitting problems. Fine-tuning with the commonly used $L^2$ regularization would help alleviate overfitting to some extent, but it can only provide an implicit inductive bias towards the pre-trained model. Therefore, we introduce a novel fine-tuning strategy that is able to produce the explicit inductive bias towards the pre-trained model by setting the model as the start point of the fine-tuning as well as a reference for the regularization. With the restriction of the reference, the capacity of the fine-tuned model will not be adapted blindly. 
Accordingly, we investigate its efficiency on different regularizers during fine-tuning and add the suffix ‘\textit{-SP}’ to the regularizers.
Assume $\mat{W}\in R^{n}$ represents the parameter matrix containing all adapted parameters of the fine-tuned model. Investigated regularizers are described as below:

$\bm{L^2}$-\textit{norm}. This is the most common penalty term used for regularization, also called weight decay. The penalty term is shown as:
\vspace{-1ex}
\begin{align}
\Theta_{2}(\mat{W})=\alpha\left\|\mat{W}\right\|_{2}^{2},
\end{align}
where $\alpha$ is the weight of the penalty.

$\bm{L^2}$\textit{-SP}. Using $\mat{W}^0$ denotes the parameter matrix of the pre-trained student model which is trained on the out-of-domain datasets. This pre-trained model is set as the start point and the reference. Thus, we penalize the $L^2$ distance between adapted parameter matrix \mat{W} and $\mat{W}^0$ to obtain:
\begin{align}
\Theta_{\textit{2-SP}}(\mat{W})=\alpha\left\|\mat{W}-\mat{W}^0\right\|_{2}^{2}
\end{align}

Considering the network architecture is usually changed when adapting the model from source datasets to a new target dataset, the penalty can be separated as two parts: one penalizes the part of the unchanged architecture $\mat{W}_{s}$ between the pre-trained and fine-tuned networks, and the second penalizes the modified architecture $\mat{W}_{m}$. As shown in Figure 1-(b), they are represented as $L_{s}$ and $L_{m}$. We obtain:
\begin{align}
\hat{\Theta}_{\textit{2-SP}}(\mat{W})=\alpha\left\|\mat{W}_{s}-\mat{W}_{s}^{0}\right\|_{2}^{2}+\beta\left\|\mat{W}_{m}\right\|_{2}^{2}
\end{align}

$\bm{L^1}$\textit{-SP}. Changing the $L^2$-norm to $L^1$-norm, we have:
\begin{align}
\hat{\Theta}_{\textit{1-SP}}(\mat{W})=\alpha\left\|\mat{W}_{s}-\mat{W}_{s}^{0}\right\|_{1}+\beta\left\|\mat{W}_{m}\right\|_{2}^{2}
\end{align}

The $L^1$ penalty encourages some parameters to be equal to corresponding parts of the pre-trained model. Consequently, $L^1$\textit{-SP} can be considered a trade-off between $L^2$\textit{-SP} and the scheme by freezing some parts of the pre-trained network.  

In our experiments, we compare the investigated regularizers based on their performance on a small target dataset. We also explore the fine-tuning performance with different network layer selections. 





\vspace{-1ex}
\section{Experiments}
\label{sec:exp}
\subsection{Dataset}

\subsubsection{In-domain target forensic dataset}
In this study, we collect a new challenging naturalistic forensic relevant dataset, called the $1^{\rm{st}}$48-UTD forensic corpus, and will release this corpus under a license in the future. The corpus is intended for forensic analysis of audio from various cities across the USA where detectives investigate actual homicides. Audio content was extracted from the USA TV program called “The First 48”, and all audio content are recorded from various real locations (e.g., interview rooms, cars, fields).  We process the raw data with the following steps: (1) extract audio at 16kHz sample rate; (2) using human manual annotation, perform diarization on the audio stream (tag speaker identity); (3) tag audio segments based on neutral-vs-stress speaker state;(4) perform automatic segmentation followed by manual check to verify segment boundaries.


Our corpus has 49 episodes, with 300 speakers, and 5041 utterances consisting of 3.5 hours of actual situational crime audio. Each episode contains disjoint speakers, and speakers in every episode are tagged as Detective, Witness, and Suspect according to their identifies.
It is a small domain-specific dataset with short utterances.  Audio consists of utterances with an average length of 2.4 s and more than 50\% of them are shorter than 2 seconds. Besides this duration constraint, context music, audio "bleeps" used for concealing harsh words, modified speech sound, and some voice-over are also contained in the audio. In this study, we use the training portion of the $1^{\rm{st}}$48-UTD corpus to fine-tune the trained student models obtained from Sec. 2.1, and evaluate its performance on the test portion. After filtering the utterances consisting of non-speech content and the speakers with fewer than three utterances, the training set consists of 3755 utterances from 228 speakers, and the test set contains 882 utterances from 39 speakers. 
\vspace{-0.5ex}
\subsubsection{Out-of-domain Voxceleb dataset}
For all experiments in Sec. 2.1, we use Voxceleb1\&2 datasets which are collected “in the wild”~\cite{nagrani2017voxceleb, chung2018voxceleb2}. The former contains 352 hours of audio from 1251 speakers, and the latter consist of 2442 hours audio from 5994 speakers. We train teacher and student speaker embedding networks on the entire Voxceleb2 dataset.
A similar data augmentation method in~\cite{snyder2018x} is adopted in the experiments.





\subsection{Experiment settings}
\textbf{T-S learning.}
For both teacher and student models, 30-dimensional log-Mel filter-banks are extracted with a frame-length of 25 ms at a 10 ms shift. Mean-normalization is applied over a sliding window of up to 3 secs, and Kaldi energy-based VAD is used to remove silence frames. 
We use the ResNet34~\cite{he2016deep} as the encoder and the learnable dictionary encoding (LDE) layer~\cite{cai2018exploring} with 64 components to aggregate frame-level features as utterance-level representations. 
The student model is initialized as identical to the teacher model at the beginning. The weights of the teacher model are fixed during student model training. With a mini-batch of 64, we use the Adam optimizer~\cite{kingma2014adam} and the learning rate is decayed using the Noam method in~\cite{vaswani2017attention}. For the teacher model, we apply a similar setting from~\cite{cai2018exploring} for network architecture and input samples. For the student, randomly cropped utterances at a fixed-length of 200 frames are utilized.

\noindent \textbf{Fine-tuning.}
With the last layer replaced, the most common way is fine-tuning only the 
last two fully-connected (FC) layers.
We also explore other two layer selections: (1) the last two FC layers plus the LDE layer and the last Residual block (Res4) of ResNet34; (2) all layers of the embedding network.
For all fine-tuning experiments, we use Adam optimizer with a learning rate decreasing by 10 in every 15 epochs. Two learning rates are used for different layers: 1e-3 used for the replaced last layer and 1e-5 used for rest of the fine-tuned layers. 
For the best results achieved, the hyper-parameters $\alpha$ and $\beta$ for $L^2$\textit{-SP} and $L^1$\textit{-SP} are set as 0.1 and 0.01, and the weight decay is set as 0.001. The back-end part comprise of LDA with dimension reduction to 200, plus centering, whitening, length normalization, and PLDA applied.


\section{Results and Analysis}

We first investigate performance of T-S learning on short utterances with different training objective functions. Table 1 presents the results of different models evaluated on our challenging $1^{\rm{st}}$48-UTD dataset. ResNet34-LDE-S represents the baseline model which is directly trained on the short crop of utterances without T-S learning. ResNet34-LDE-L is the teacher model trained on the long utterances (3s-8s). 
As expected, the baselines produce better results than the teacher models by decreasing the EERs from 14.97\% and 13.83\% to 13.35\% and 12.79\% using softmax and A-softmax respectively. The results show that training the speaker embedding network with short utterances can improve its performance 
on short evaluation utterances.
However, the baseline cannot obtain further improvement without using the knowledge transferred from the teacher model.

\begin{table}[bhbp]
  \renewcommand{\arraystretch}{0.95}

  \caption{Performance of baseline systems, teacher and student models on the $1^{\rm{st}}$48-UTD dataset. Column "Distillation" represents the objective functions used for student model training.}
  \vspace{-3.2mm}
  \begin{tabular}{ccc}
  \bottomrule
  System              & \multicolumn{1}{c}{Distillation} & \multicolumn{1}{c}{EER(\%)} \\ \midrule
  \multicolumn{3}{c}{(a) $L_{\mathrm{class}}$ = Softmax}                                  \\ \midrule
  \multicolumn{1}{l}{ResNet34-LDE-L} & \multicolumn{1}{c}{-} & 14.97   \\ 
  \multicolumn{1}{l}{ResNet34-LDE-S} & \multicolumn{1}{c}{-} & 13.35   \\ \hline
  \multirow{3}{*}{ResNet34-LDE}        & $L_{\mathrm{class}}$+$L_{\mathrm{KLD}}$            & 12.31   \\ 
                                      & $L_{\mathrm{class}}$+$L_{\mathrm{EMD}}$            & 12.03   \\ 
                                      & $L_{\mathrm{class}}$+$L_{\mathrm{KLD}}$+$L_{\mathrm{EMD}}$       & \textbf{11.65}   \\ \midrule \midrule
  \multicolumn{3}{c}{(b) $L_{\mathrm{class}}$ = A-Softmax}                                \\ \midrule
  \multicolumn{1}{l}{ResNet34-LDE-L} & \multicolumn{1}{c}{-} & 13.83   \\ 
  \multicolumn{1}{l}{ResNet34-LDE-S} & \multicolumn{1}{c}{-} & 12.79   \\ \hline
  \multirow{3}{*}{ResNet34-LDE}        & $L_{\mathrm{class}}$+$L_{\mathrm{KLD}}$           & 11.85   \\ 
                                      & $L_{\mathrm{class}}$+$L_{\mathrm{EMD}}$          & 11.34   \\ 
                                      & $L_{\mathrm{class}}$+$L_{\mathrm{KLD}}$+$L_{\mathrm{EMD}}$       & \textbf{11.12}   \\ \bottomrule
  \end{tabular}
\end{table}

With the same embedding network utilized, the student model can still boost the performance with the advantage of the proposed objective function for T-S learning. As shown in Table 1, different student objective functions contribute to different levels of performance improvement, and all student models outperform their teacher and baseline models. In Table 1-(a), when using the softmax, three student models reduce the EER of the baseline system to 12.31\%, 12.03\%, and 11.65\%, respectively. While using the A-softmax loss can still further boost the performance to 11.85\%, 11.34\%, and 11.12\%. Based on the results, we find that the models optimizted by {$L_{\mathrm{class}}+L_{\mathrm{EMD}}$} consistently outperform that using {$L_{\mathrm{class}}+L_{\mathrm{KLD}}$}. It indicates that the speaker embedding quality is more relevant to system performance. Compared with baselines, it is consistent to observe that models optimized by {$L_{\mathrm{class}}+L_{\mathrm{KLD}}+L_{\mathrm{EMD}}$} can achieve the best performances for both softmax and A-softmax with the highest reductions of EER by 12.7\% and 13.1\% relatively. This makes sense since the student model can learn additional speaker embedding knowledge from long utterances besides the speaker discriminative power, and it is more effective to compensate for short utterances.


Considering two student models optimized by the proposed objective function  {$L_{\mathrm{class}}+L_{\mathrm{KLD}}+L_{\mathrm{EMD}}$} and {$L_{\mathrm{class}}+L_{\mathrm{EMD}}$} with A-softmax used, they produce the best two performances on 2-second segments of Voxceleb1 evaluation set with EERs of 6.38\% and 6.51\%, respectively. Meanwhile, they also produce the best two performances on the target dataset in Table 1. Thus, we pick them as the start points for the follow-up fine-tuning step. 
\begin{table}[t]
\caption{Performance comparison of different regularization strategies and layer selections on the best two student models. The first column represents proposed regularizers, the first row represents the different layer selections.}
\vspace {-3.2mm}
\hspace{0.1mm}
\begin{tabular}{cccc}
\bottomrule
\multicolumn{1}{c}{}                 & \multicolumn{1}{c}{Last 2 FC} & \multicolumn{1}{c}{\begin{tabular}[c]{@{}c@{}}Last 2 FC+LDE\\+Res4\end{tabular}} & \multicolumn{1}{c}{All Layers} \\ \hline
\multicolumn{4}{c}{(a) A-Softmax \& $L_{\mathrm{class}}$+$L_{\mathrm{KLD}}$+$L_{\mathrm{EMD}}$}                                                                                                                                             \\ \hline
{\color[HTML]{000000} w/o regularizer} & {\color[HTML]{000000} 11.03}   & {\color[HTML]{000000} 10.88}                                                        & {\color[HTML]{000000} 11.36}    \\
{\color[HTML]{000000} $L^2$\textit{-norm}}         & {\color[HTML]{000000} 10.85}   & {\color[HTML]{000000} 10.52}                                                        & {\color[HTML]{000000} 10.97}    \\
{\color[HTML]{000000} $L^1$\textit{-SP}}           & {\color[HTML]{000000} 10.67}   & {\color[HTML]{000000} 10.30}                                                        & {\color[HTML]{000000} 10.14}    \\
$L^2$\textit{-SP}                                  & {\color[HTML]{000000} \textbf{10.48}}   & {\color[HTML]{000000} \textbf{10.13}}                                                        & {\color[HTML]{000000} \textbf{9.85}}     \\ \hline \hline
\multicolumn{4}{c}{(b) A-Softmax \& $L_{\mathrm{class}}$+$L_{\mathrm{EMD}}$}                                                                                                                                                  \\ \hline
w/o regularizer                        & {\color[HTML]{000000} 11.28}   & {\color[HTML]{000000} 11.20}                                                        & {\color[HTML]{000000} 11.52}    \\
$L^2$\textit{-norm}                                & {\color[HTML]{000000} 11.21}   & {\color[HTML]{000000} 11.03}                                                        & {\color[HTML]{000000} 11.29}    \\
$L^1$\textit{-SP}                                  & {\color[HTML]{000000} 10.92}   & {\color[HTML]{000000} 10.53}                                                        & {\color[HTML]{000000} 10.25}    \\
$L^2$\textit{-SP}                                  & {\color[HTML]{000000} \textbf{10.70}}   & {\color[HTML]{000000} \textbf{10.31}}                                                        & {\color[HTML]{000000} \textbf{10.09}} \\ \bottomrule
\end{tabular}
\end{table}

Table 2 shows fine-tuning results on the $1^{\rm{st}}$48-UTD dataset. We find that adding the regularizer improves fine-tuning performance. In Table 2-(a), the regularizers $L^2$\textit{-norm}, $L^1$\textit{-SP}, and $L^2$\textit{-SP} achieve their best performances with 10.52\%, 10.14\%, and 9.85\% in EER.
In the lower Table 2-(b), they achieve their best performances with EERs of 11.03\%, 10.25\%, and 10.09\%, respectively. We observe the trend that with the restriction of reference, $L^2$\textit{-SP} and $L^1$\textit{-SP} consistently outperform the $L^2$\textit{-norm} for all three layer selection methods. Thus, it is beneficial to fine-tune the pre-trained model with explicit inductive bias, especially when the target dataset is small. For $L^2$\textit{-norm}, the problem of overfitting to a small amount of data will be more severe when lower layers are adapted, and it is shown that performance of fine-tuning with it begins to degrade from adapting part of the network to adapting all layers.
Comparing to the best results achieved by $L^1$\textit{-SP} and $L^2$\textit{-norm} in Table 2, $L^2$\textit{-SP} outperforms them and reduces EERs by relative 2.9\% and 6.4\%, respectively. Considering the small size, short duration and other variations mentioned in Sec. 3.1.1 for the $1^{\rm{st}}$48-UTD dataset, this fine-tuning strategy not only helps resist forgetting the features learned from large number of speakers in the source domain, but also adapts the model to the target domain efficiently. Our results suggest that the $L^2$\textit{-SP} and $L^1$\textit{-SP} are definitely more efficient than weight decay especially when lower layers are adapted, and $L^2$\textit{-SP} can provide more strict regularization compared to $L^1$\textit{-SP}.

\section{Conclusion}
In this study, we focused on approaches for short utterance speaker verification on a small challenging naturalistic forensic dataset. Our results indicated that speaker discriminative power and embedding similarity are two significant points for short utterance speaker verification. The proposed objective function for teacher-student learning was shown to transfer critical knowledge of these two points from long utterances to short utterances more effectively. Using our novel fine-tuning strategy, speaker embedding networks were adapted to a new small target dataset while preserving the speaker discriminative power learned from large number of speakers. Experiment results show that our approaches can significantly improve performance of speaker verification systems on small domain-specific short utterance datasets.


\vfill\pagebreak

\newpage

\bibliographystyle{IEEEtran}
\bibliography{is2019_sed.bib}

\begin{thebibliography}{10}
\providecommand{\url}[1]{#1}
\csname url@samestyle\endcsname
\providecommand{\newblock}{\relax}
\providecommand{\bibinfo}[2]{#2}
\providecommand{\BIBentrySTDinterwordspacing}{\spaceskip=0pt\relax}
\providecommand{\BIBentryALTinterwordstretchfactor}{4}
\providecommand{\BIBentryALTinterwordspacing}{\spaceskip=\fontdimen2\font plus
\BIBentryALTinterwordstretchfactor\fontdimen3\font minus
  \fontdimen4\font\relax}
\providecommand{\BIBforeignlanguage}[2]{{%
\expandafter\ifx\csname l@#1\endcsname\relax
\typeout{** WARNING: IEEEtran.bst: No hyphenation pattern has been}%
\typeout{** loaded for the language `#1'. Using the pattern for}%
\typeout{** the default language instead.}%
\else
\language=\csname l@#1\endcsname
\fi
#2}}
\providecommand{\BIBdecl}{\relax}
\BIBdecl

\bibitem{snyder2016deep}
D.~Snyder, P.~Ghahremani, D.~Povey, D.~Garcia-Romero, Y.~Carmiel, and
  S.~Khudanpur, ``Deep neural network-based speaker embeddings for end-to-end
  speaker verification,'' in \emph{2016 IEEE Spoken Language Technology
  Workshop (SLT)}.\hskip 1em plus 0.5em minus 0.4em\relax IEEE, 2016, pp.
  165--170.

\bibitem{zhang2017end}
C.~Zhang and K.~Koishida, ``End-to-end text-independent speaker verification
  with triplet loss on short utterances.'' in \emph{Interspeech}, 2017, pp.
  1487--1491.

\bibitem{li2017deep}
C.~Li, X.~Ma, B.~Jiang, X.~Li, X.~Zhang, X.~Liu, Y.~Cao, A.~Kannan, and Z.~Zhu,
  ``Deep speaker: an end-to-end neural speaker embedding system,'' \emph{arXiv
  preprint arXiv:1705.02304}, 2017.

\bibitem{wan2018generalized}
L.~Wan, Q.~Wang, A.~Papir, and I.~L. Moreno, ``Generalized end-to-end loss for
  speaker verification,'' in \emph{2018 IEEE International Conference on
  Acoustics, Speech and Signal Processing (ICASSP)}.\hskip 1em plus 0.5em minus
  0.4em\relax IEEE, 2018, pp. 4879--4883.

\bibitem{snyder2018x}
D.~Snyder, D.~Garcia-Romero, G.~Sell, D.~Povey, and S.~Khudanpur, ``X-vectors:
  Robust dnn embeddings for speaker recognition,'' in \emph{2018 IEEE
  International Conference on Acoustics, Speech and Signal Processing
  (ICASSP)}.\hskip 1em plus 0.5em minus 0.4em\relax IEEE, 2018, pp. 5329--5333.

\bibitem{xia2019cross}
W.~Xia, J.~Huang, and J.~H. Hansen, ``Cross-lingual text-independent speaker
  verification using unsupervised adversarial discriminative domain
  adaptation,'' in \emph{IEEE International Conference on Acoustics, Speech and
  Signal Processing (ICASSP)}.\hskip 1em plus 0.5em minus 0.4em\relax IEEE,
  2019, pp. 5816--5820.

\bibitem{dehak2010front}
N.~Dehak, P.~J. Kenny, R.~Dehak, P.~Dumouchel, and P.~Ouellet, ``Front-end
  factor analysis for speaker verification,'' \emph{IEEE Transactions on Audio,
  Speech, and Language Processing}, vol.~19, no.~4, pp. 788--798, 2010.

\bibitem{kanagasundaram2011vector}
A.~Kanagasundaram, R.~Vogt, D.~B. Dean, S.~Sridharan, and M.~W. Mason,
  ``I-vector based speaker recognition on short utterances,'' in
  \emph{Proceedings of the 12th Annual Conference of the International Speech
  Communication Association}.\hskip 1em plus 0.5em minus 0.4em\relax
  International Speech Communication Association (ISCA), 2011, pp. 2341--2344.

\bibitem{xie2019utterance}
W.~Xie, A.~Nagrani, J.~S. Chung, and A.~Zisserman, ``Utterance-level
  aggregation for speaker recognition in the wild,'' in \emph{ICASSP 2019-2019
  IEEE International Conference on Acoustics, Speech and Signal Processing
  (ICASSP)}.\hskip 1em plus 0.5em minus 0.4em\relax IEEE, 2019, pp. 5791--5795.

\bibitem{jung2019short}
J.-w. Jung, H.-S. Heo, H.-j. Shim, and H.-J. Yu, ``Short utterance compensation
  in speaker verification via cosine-based teacher-student learning of speaker
  embeddings,'' in \emph{2019 IEEE Automatic Speech Recognition and
  Understanding Workshop (ASRU)}.\hskip 1em plus 0.5em minus 0.4em\relax IEEE,
  2019, pp. 335--341.

\bibitem{hajavi2019deep}
A.~Hajavi and A.~Etemad, ``A deep neural network for short-segment speaker
  recognition,'' \emph{arXiv preprint arXiv:1907.10420}, 2019.

\bibitem{gusev2020deep}
A.~Gusev, V.~Volokhov, T.~Andzhukaev, S.~Novoselov, G.~Lavrentyeva, M.~Volkova,
  A.~Gazizullina, A.~Shulipa, A.~Gorlanov, A.~Avdeeva \emph{et~al.}, ``Deep
  speaker embeddings for far-field speaker recognition on short utterances,''
  \emph{arXiv preprint arXiv:2002.06033}, 2020.

\bibitem{mandasari2011evaluation}
M.~I. Mandasari, M.~McLaren, and D.~A. van Leeuwen, ``Evaluation of i-vector
  speaker recognition systems for forensic application,'' 2011.

\bibitem{poddar2017speaker}
A.~Poddar, M.~Sahidullah, and G.~Saha, ``Speaker verification with short
  utterances: a review of challenges, trends and opportunities,'' \emph{IET
  Biometrics}, vol.~7, no.~2, pp. 91--101, 2017.

\bibitem{al2017enhanced}
A.~K.~H. Al-Ali, D.~Dean, B.~Senadji, V.~Chandran, and G.~R. Naik, ``Enhanced
  forensic speaker verification using a combination of dwt and mfcc feature
  warping in the presence of noise and reverberation conditions,'' \emph{IEEE
  Access}, vol.~5, pp. 15\,400--15\,413, 2017.

\bibitem{machado2019forensic}
T.~J. Machado, J.~Vieira~Filho, and M.~A. de~Oliveira, ``Forensic speaker
  verification using ordinary least squares,'' \emph{Sensors}, vol.~19, no.~20,
  p. 4385, 2019.

\bibitem{hinton2015distilling}
G.~Hinton, O.~Vinyals, and J.~Dean, ``Distilling the knowledge in a neural
  network,'' \emph{arXiv preprint arXiv:1503.02531}, 2015.

\bibitem{liu2017sphereface}
W.~Liu, Y.~Wen, Z.~Yu, M.~Li, B.~Raj, and L.~Song, ``Sphereface: Deep
  hypersphere embedding for face recognition,'' in \emph{Proceedings of the
  IEEE conference on computer vision and pattern recognition}, 2017, pp.
  212--220.

\bibitem{lu2017knowledge}
L.~Lu, M.~Guo, and S.~Renals, ``Knowledge distillation for small-footprint
  highway networks,'' in \emph{2017 IEEE International Conference on Acoustics,
  Speech and Signal Processing (ICASSP)}.\hskip 1em plus 0.5em minus
  0.4em\relax IEEE, 2017, pp. 4820--4824.

\bibitem{wang2019knowledge}
S.~Wang, Y.~Yang, T.~Wang, Y.~Qian, and K.~Yu, ``Knowledge distillation for
  small foot-print deep speaker embedding,'' in \emph{ICASSP 2019-2019 IEEE
  International Conference on Acoustics, Speech and Signal Processing
  (ICASSP)}.\hskip 1em plus 0.5em minus 0.4em\relax IEEE, 2019, pp. 6021--6025.

\bibitem{bimbot2004tutorial}
F.~Bimbot, J.-F. Bonastre, C.~Fredouille, G.~Gravier, I.~Magrin-Chagnolleau,
  S.~Meignier, T.~Merlin, J.~Ortega-Garc{\'\i}a, D.~Petrovska-Delacr{\'e}taz,
  and D.~A. Reynolds, ``A tutorial on text-independent speaker verification,''
  \emph{EURASIP Journal on Advances in Signal Processing}, vol. 2004, no.~4, p.
  101962, 2004.

\bibitem{nagrani2017voxceleb}
A.~Nagrani, J.~S. Chung, and A.~Zisserman, ``Voxceleb: a large-scale speaker
  identification dataset,'' \emph{arXiv preprint arXiv:1706.08612}, 2017.

\bibitem{chung2018voxceleb2}
J.~S. Chung, A.~Nagrani, and A.~Zisserman, ``Voxceleb2: Deep speaker
  recognition,'' \emph{arXiv preprint arXiv:1806.05622}, 2018.

\bibitem{he2016deep}
K.~He, X.~Zhang, S.~Ren, and J.~Sun, ``Deep residual learning for image
  recognition,'' in \emph{Proceedings of the IEEE conference on computer vision
  and pattern recognition}, 2016, pp. 770--778.

\bibitem{cai2018exploring}
W.~Cai, J.~Chen, and M.~Li, ``Exploring the encoding layer and loss function in
  end-to-end speaker and language recognition system,'' \emph{arXiv preprint
  arXiv:1804.05160}, 2018.

\bibitem{kingma2014adam}
D.~P. Kingma and J.~Ba, ``Adam: A method for stochastic optimization,''
  \emph{arXiv preprint arXiv:1412.6980}, 2014.

\bibitem{vaswani2017attention}
A.~Vaswani, N.~Shazeer, N.~Parmar, J.~Uszkoreit, L.~Jones, A.~N. Gomez,
  {\L}.~Kaiser, and I.~Polosukhin, ``Attention is all you need,'' in
  \emph{Advances in neural information processing systems}, 2017, pp.
  5998--6008.

\end{thebibliography}

\end{document}